\documentstyle[epsfig,11pt]{article}
\begin{document}
%
%

\begin{titlepage}

\pagenumbering{arabic}
\vspace*{-1.5cm}
\begin{tabular*}{15.cm}{l@{\extracolsep{\fill}}r}
2 October, 1997                                        
&
                                                         IReS 97-29  \\
&                                                     PRA-HEP 97/16
\\
&\\ \hline
\end{tabular*}
\vspace*{2.cm}
\begin{center}
\Large 

{\bf
 Simulation of Bose-Einstein effect \\
   using space-time aspects of Lund string fragmentation model
} \\
\vspace*{2.cm}
\normalsize {

   {\bf \v{S}\'arka Todorova-Nov\'{a}} \\
\vspace*{0.5cm}
   {\footnotesize Nuclear Centre, Faculty of Mathematics and Physics,  
   Charles University, Prague \\
             and IReS Strasbourg \footnote{e-mail: nova@sbgaxp.in2p3.fr}} \\
\vspace*{0.5cm}

\vspace*{0.5cm}

   {\bf  Ji\v{r}\'{\i} Rame\v{s} } \\
\vspace*{0.5cm}
   {\footnotesize Institute of Physics, Academy of Sciences
   of the Czech Republic, Prague \footnote{e-mail: rames@fzu.cz}} \\ 

}
\end{center}

\begin{abstract}
      The experimentally observed enhancement of number of close boson
 pairs in $e^{+}e^{-}$ collisions is reproduced by local weighting
 according to the quantum mechanical prescriptions for
 production of identical bosons. The space-time picture
 of the process, inherently present in the Lund fragmentation model,
 is explicitly used.

 The model is used to check
 systematic errors in the W mass measurements due to the Bose-Einstein
 effect.

 The possibility of direct implementation of the Bose-Einstein effect
 into string fragmentation is discussed.
\end{abstract}

\end{titlepage}
\newpage


\pagestyle{plain}

\setcounter{page}{1}



\section{Introduction}

   Recently, the Bose-Einstein (BE) effect in particle production
 in $e^{+}e^{-}$ annihilations received particular attention as
 LEP doubled its collision energy, allowing for direct production
 of WW pairs.

  The  influence of the BE effect on the measured W mass at LEP2 was first
 investigated in Ref.\cite{LS}. The standard JETSET implementation of BE effect
 (routine LUBOEI, Ref.\cite{JETSET}), used in this study,  reshuffles  momenta
 of generated particles to increase the fraction of close boson pairs
 according to a phenomenological parameterization. The method has some
 technical shortcomings (as local violation of
 energy/momentum conservation laws)
 but the basic problem is that it actually doesn't make any connection
 between the quantum mechanical (QM) origin of the effect and 
 its observable consequences, and
 therefore
 it has relatively low predictive power; as a result,
  only a very vague  estimation of systematic error could be
 drawn out \cite{YR}. Recently, other studies \cite{RM},\cite{JZ}
 used the phenomenological
 formula for global event weighting to extract the systematic uncertainty
 on the W mass measurement;
 this uncertainty was found below  20 \cite{RM} or 30 \cite{JZ} MeV,
 but the procedure backfired by predicting a change in $R_b$ and
 $R_c$ of 10-20 \% in $Z^0$ decay \cite{RM}, which is not observed.

      The problem with all studies mentioned above
 is that they are based uniquely on
 the single external appearance of the BE effect -- the enhancement of
 production of close pairs of identical bosons -- while this is probably
 only the most visible  consequence of more fundamental physical processes
 taking part in the hadronization.

  A fairly better way towards understanding the BE interference
 consists in implementing it into the simulation
 starting from "first principles", i.e. starting from QM formulae,
 and only then to check the consistency of the
 predictions with experimental data.

   While the possibility to include QM interference effects into string
 fragmentation  models (Artru-Mennessier, LUND) was pointed out
 a long time ago \cite{AH},\cite{AB}, only quite recently
 a Monte-Carlo (MC) implementation of these ideas  appeared \cite{markus}.

   The method presented in this paper
 -- while having some common features with the LUND approach --
 simplifies the full QM treatment by resigning on higher order correlations;
 also, the global event weighting is replaced by ``local'' implementation
 of BE correlations.

   The behaviour of simulated data is discussed, and they are compared to the
 experimental data. An alternative tool for study of particle correlations
 -- factorial moments -- is used to compare the standard JETSET simulation
 (LUBOEI) with the new one, presented in this paper.
 
   The influence of the BE effect on the measurement of the W mass is
 investigated, and the systematic uncertainty due to this effect is
 estimated. 
  
  The last section of this paper deals with  possible strategies
 and simplifications for future simulations of the BE effect.

\section{Correlation function}

  The Bose-Einstein interference (or Hanbury-Brown-Twiss effect  in
  astronomy) is experimentally seen as an enhanced probability of
  observing two (and more) identical bosons with a similar momentum.
    In the language of QM, this enhanced probability arises from the
 symmetrization of the amplitude with respect to the exchange of
 identical bosons.

  If we describe the one-particle wave function  by a planar wave

 \begin{center} {

 $  \Phi_{i} \sim \exp{\{\frac{-i}{\hbar} p.(x-x_i)\}}  $

  } \end{center}

 where $p$ is the 4-momentum and $x_i$ the production vertex of the particle,
then the symmetrization of an $N$ particle wave function 
 $ \Phi_1 \Phi_2 ... \Phi_N $
 in the case of $N$ identical bosons gives the amplitude:

\begin{equation}
\Psi(N) = \frac{1}{\sqrt{N!}} \sum_{i_{i}} \exp{\{ \frac{-i}{\hbar} 
  [ p_{i1}(x-x_1) +p_{i2}(x-x_2)+     ... +p_{iN}(x-x_N) ]\} }
\end{equation}

 and the probability:

\begin{eqnarray}
 P(N) & = & |\Psi(N)|^2 =  \nonumber   \\
      & = & \frac{1}{N!} \sum_{i_{i},j_{i}} \exp{\{ \frac{-i}{\hbar}
       [ (p_{i1}-p_{j1})x_1+(p_{i2}-p_{j2})x_2+
     ... +(p_{iN}-p_{jN})x_N ] \} }  \nonumber   \\
      & = & 1 + \underbrace { \frac{1}{N!} \sum_{i\neq j,k\neq l}
                            \exp{\{\frac{-i}{\hbar}
   [(p_{i}-p_{j})x_{k}+(p_{j}-p_{i})x_{l})]\} } }_{2-part.correlations} 
               \nonumber  \\
      &   & +  \underbrace {\frac{1}{N!} \sum_{i\neq j\neq m,k\neq l\neq n}
         \exp{\{ \frac{-i}{\hbar} [(p_{i}-p_{j})x_{k}+(p_{j}-p_{m})x_{l}+
        (p_{m}-p_{i})x_{n}) ]\} } }_{3-part.correlations}     \\
      &   & + \cdots   \nonumber
\end{eqnarray}

   The number of terms in the symmetrised formula increases as  $(N!)^2$ 
(which
 already indicates the complexity of evaluating higher orders with
 many identical bosons).

 Throughout this paper, only 2- and 3- particle interference
 terms will be used, rewritten in the convenient form:

\begin{itemize}
\item  2-particle correlations $ \sim \sum_{i<j,k\neq l}   \
   cos[(p_{i}-p_{j})\cdot (x_{k}-x_{l})/\hbar] $
\item  3-particle correlations $ \sim
      2 \sum_{i<j\neq m,k\neq l\neq n}  \
        cos\{[(p_{i}-p_{j})(x_{k}-x_{n})+(p_{j}-p_{m})(x_{l}-x_{n})]/\hbar\} 
                                                $

\end{itemize}

  In general, all interference terms can be expressed  in terms of
 $dp\cdot dx$, the invariant product of the difference in momentum and
 in space-time
 distance of the production vertices. How this variable can be evaluated
 in the string fragmentation model is discussed in the next section.

\section{Space-time picture of string fragmentation}

  The Lund string fragmentation model \cite{LUN83}, in the form of its MC
 implementation JETSET \cite{JETSET} is commonly used in simulations of
 hadronic final states at high energies due to its ability
 to reproduce the experimental data quite well.
   The interesting property of this model with respect to the study
 of the BE effect lies in the possibility to reconstruct the space-time
 picture of string breaking. The initial position of string
 fragment -- the production point of final hadron -- can then be derived.

\vspace{-2.cm}

 \begin{figure}[bth]
\mbox{\epsfig{file=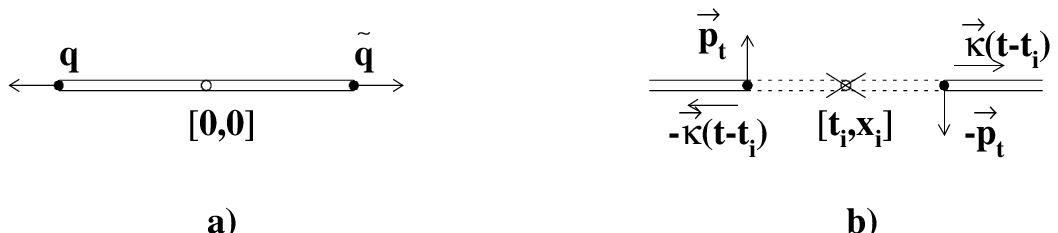,height=5cm,width=15cm}}
\parbox{15.cm}{\caption{\protect\small \sl
}
\label{fig:string}}
\end{figure}

  Schematically, the situation is shown in  Fig.\ref{fig:string}a); the
 original string, spanned between two endpoint partons, carries in
 its rest frame a longitudinal energy density
 $ |\vec{\kappa}| \simeq 1$ GeV/fm   ($ \vec{\kappa}$ is called "string
 tension").
 The string breaks by creating a new quark-antiquark pair
 (``tunneling'' mechanism); the new quarks are supposed to be produced with
 a zero longitudinal initial momentum ( longitudinal with respect to the string
 direction) and a non-zero transverse momentum ($ \pm \vec{p}_t $);
 due to the string tension, they separate and move in opposite directions,
 acquiring a momentum  $ p_{long}=\pm |\vec{\kappa}| dt
 $(Fig:~\ref{fig:string}b).

\vspace{-2.cm}

 \begin{figure}[bth]
\mbox{\epsfig{file=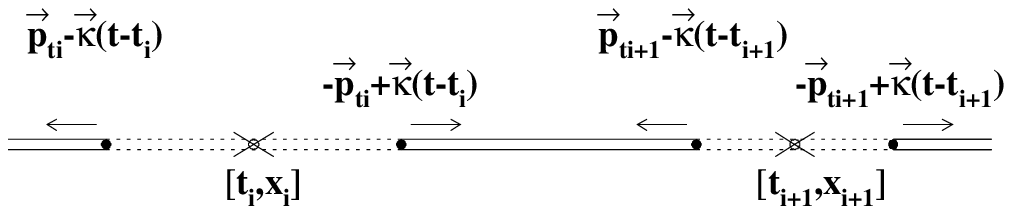,height=5cm,width=15cm}}
\parbox{15.cm}{\caption{ }
\label{fig:str2}}
\end{figure}

   Two neighbour string breakings give birth to a hadron ; its energy and
 momentum can be expressed in terms of space-time coordinates of the string
 breaking (see Fig.\ref{fig:str2})
 \begin{eqnarray}
     E_{had}       & = & \kappa dl = \kappa  |x_i-x_{i+1}|  \nonumber \\
    \vec{p}_{had}  & = & \vec{p}_{t_i} + \vec{p}_{t_{i+1}} + \vec{\kappa}
    (t_i-t_{i+1}) 
\label{eq:f1}
\end{eqnarray}

  Alternatively, the coordinates of string breaking can be expressed as
 a function of the momenta of final hadrons. Each breaking divide the total
 number of final hadrons into two parts -- left [L] and right [R]-- according
 to the part of string they came from.
 Presuming that the string starts
 to expand from point [0,0] in its rest frame, then the coordinates of
 the $i$-th breaking are: 
\begin{eqnarray}
     x_{i} & = &( \sum_{L_i} E_{had} - \sum_{R_i} E_{had})/\kappa \nonumber \\
     t_{i} & = &( p_{0} - \sum_{L_i} p_{had_{long}})/\kappa   
\label{eq:fx}    
\end{eqnarray}
  where $p_0$ stands for the initial momentum of the endpoint partons.
   Therefore, the calculation of coordinates of string breaking is
 straightforward  for a simple $q\tilde{q}$ string.

   However, things become considerably more complicated in the case of
 gluon radiation because of the complicated string movement around gluon
 corners (kinks). The algorithm finding the position of the string 
 at the moment of its breaking is actually the most complicated part of
 the whole
 simulation of the BE effect. It follows closely the fragmentation process 
 in JETSET and evaluates the space-time coordinates in parallel with
 the generation of hadron' momenta.

  Once the points where a string broke are found, the production vertices
 of hadrons can be calculated. A kind of convention needs to be adopted here,
 because the hadron is not a point-like object and because the two endpoint 
 string
 breakings are causally disconnected. Therefore, by the production vertex
 of the hadron we will understand the barycentre of the string piece forming
 the hadron in the frame where the two endpoint breakings occur simultaneously.
   For a simple $q\tilde{q}$ string in its rest frame, the  coordinates of
 the production vertex of the hadron will be: 
\begin{equation}
    \vec{x}_{had} = 0.5 ( \vec{x}_i + \vec{x}_{i+1} )  ; \hspace{0.5cm}
    t_{had} = 0.5 ( t_i + t_{i+1} ) 
\label{eq:f2}
\end{equation}

     Since  one is usually only interested
 in the momentum spectrum of the produced hadrons, the space-time history
 of the fragmentation is not evaluated in JETSET. Therefore, this
 information had to be traced back and added into the standard event record.

\begin{figure}[bth]
\mbox{\epsfig{file=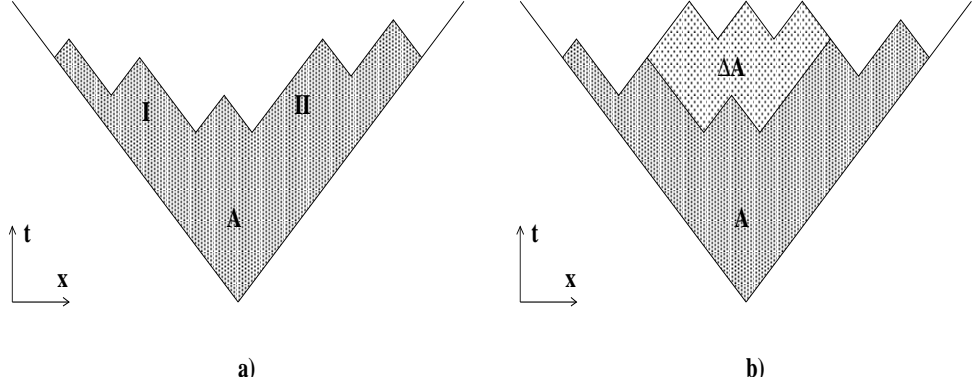,height=5.5cm,width=15cm}}
\parbox{15.cm}{\caption{\protect\small \sl
   Fig.~\ref{fig:area}a) shows the space-time diagram of string fragmentation.
 The shaded area A is the area spanned by the string. Fig.~\ref{fig:area}b)
 shows the string area difference $\Delta A$ corresponding to the exchange
 of hadrons I and II. }
\label{fig:area}}
\end{figure}

  Knowing  the space-time distribution of the hadrons, 
 we are now able to evaluate  the $ dp \cdot dx $ terms in
 the correlation function of section 2. The problem is that for
  the moment, our correlation function (Eq.2) does not take into
  account the dynamics of the process of hadronization. We can however use
 the QM framework of the Lund fragmentation model developed in
 Ref.\cite{AH},\cite{AB},\cite{markus}. On the basis of
 the argumentation provided in these studies, not only the probability
 of string breaking can be related to the area $A$ spanned by the string  
 (the space-time integral over string movement, Fig.\ref{fig:area})
 but also the phase
 of the amplitude, so that the amplitude of the string fragmentation process
 can be written as
\begin{equation}
   M = \ exp ( i\kappa - b/2 ) A
\label{eq:amp}
\end{equation}
  where $b$ is a parameter tuned to the experimental data.

  The symmetrization of this amplitude with respect to the exchange
 of $N$ identical bosons yields
\begin{equation}
  M \rightarrow M_{sym} =\frac{1}{\sqrt{N!}} \sum_{i=1}^{N!}
             \   exp ( i\kappa - b/2 ) A_i
\end{equation}
 and the amplitude squared  can be written as
\begin{eqnarray}
 |M_{sym}|^2 & = &  \frac{1}{N!} \sum_{i,j=1}^{N!}
                   \  exp[ i \kappa ( A_i - A_j ) ]
                                \   exp [ \frac{-b}{2} ( A_i + A_j ) ]
                    \nonumber \\
             & = &  \frac{1}{N!} \{ \sum_{j=1}^{N!} \ exp (- b A_j) +
                     \nonumber \\
             &   &  \sum_{i,j,A_i\geq A_j} 2 \ cos [ \kappa (A_i-A_j) ]
                  \ exp [ \frac{-b}{2} ( A_i - A_j ) ] \ exp (- b A_j ) \}
                    \nonumber  \\
             & = &  \frac{1}{N!} \sum_{j=1}^{N!} \ exp (- b A_j)
                    \{ 1 + \sum_{i,A_i\geq A_j} 2 \ cos ( \kappa \Delta A_{ij} )
                        \  exp ( \frac{-b}{2} \Delta A_{ij} )  \} 
\label{eq:sup}
\end{eqnarray}

  The interference appears in the formula as an additional weight
 depending only on the string area difference. This difference is
 shown in Fig.~\ref{fig:area}b) for the exchange of two hadrons (I,II).
   It can be shown (see Appendix A) that this area difference
 (times $\kappa$)
 is equal to the $ dp \cdot dx $ term :

\begin{equation}
     \kappa \Delta A = dp \cdot dx
\label{eq:supa}
\end{equation}

\begin{figure}[bth]
\mbox{\epsfig{file=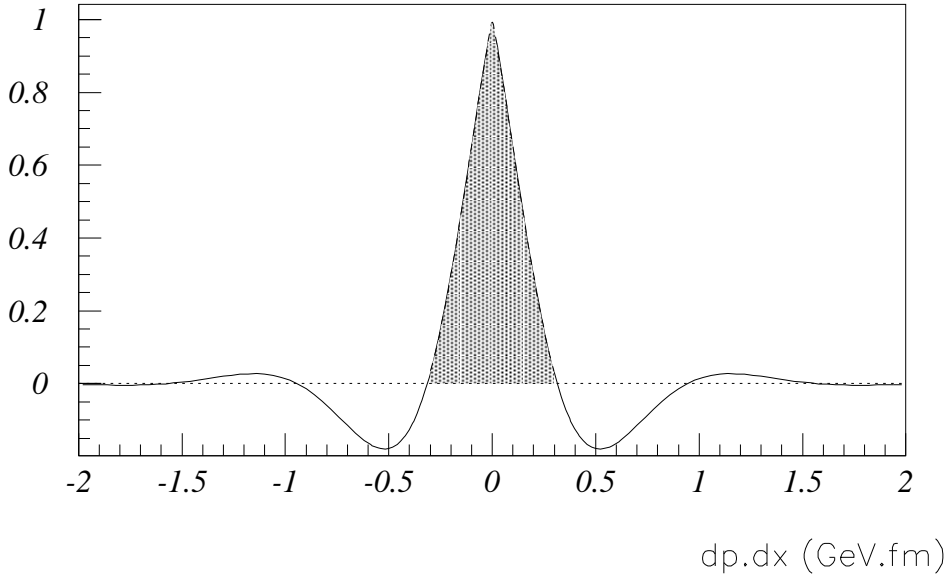,height=10cm,width=15cm}}
\parbox{15.cm}{\caption{\protect\small \sl
 The shape of the interference term in Eq.8.
}
\label{fig:corfun}}
\end{figure}

  The comparison of Eq.2 to Eq.\ref{eq:sup} shows that the simple correlation 
 function is now damped
  by an exponential term (see Fig.~\ref{fig:corfun}). The effect
 is  concentrated in a small region around the origin of the $ dp \cdot dx $
 distribution; this is where the close pairs ( or multiplets) are 
 expected to be located.

\section{Simulation strategy }

   Formula \ref{eq:sup} provides a recipe for how to include
 all interference effects into the simulation via global event weights.
 However, the evaluation of all interference terms for all possible
 boson exchanges remains quite complicated;
 this is the way the Bose-Einstein
 effect is handled in \cite{markus}.

   As already mentioned above, we have chosen a simplified way to
 implement the BE interference. This works only with 2-particle,
 eventually 3-particle, exchanges. Without higher order interference terms,
  formula \ref{eq:sup} can hardly be used as such since one cannot achieve
 a proper normalization nor handle safely negative weights. On the other hand,
 we know that the effect is very localised in the configuration space
 and that the observed
 enhancement in the production of close boson pairs is due to the peak
 in the $ dp\cdot dx $ distribution; therefore, the generated events must
 contain pairs of bosons for which the products $ dp\cdot dx $ lie 
 in the interval within
 the shaded area of Fig.~\ref{fig:corfun}. 
   
    The simulation program was built from the beginning on this
 {\em qualitative} feature of the BE interference, and  
 several simplifications were therefore introduced 
 in order to have the possibility
 to study various aspects of the production of close boson pairs. On the
 level of the correlation function -- built from 2- and 3- particle interference
 terms only -- we omit the secondary peaks
 and minima of interference terms, setting their minimal value to 0.
 This allows us to {\em force} the production of close boson pairs, because
 all configurations outside the central peak are rejected. We keep
 this simplified form of the interference term throughout this paper because
 it provides results which are in a good agreement with experimental data.
 The form of the interference term can be easily changed in the simulation
 program, and the dependence of the result on its modifications can be
 studied. 

\vspace{0.5cm}

\begin{table}[bth]

\begin{minipage}[t]{8.cm}
\begin{tabular}{|c|c|}
\hline
   Particle type & Production rate \\
                 & ( LEP \cite{tun}) \\
\hline
   $\pi^{\pm} $ & $ 17.1 \pm 0.4 $ \\
\hline
   $\pi^{0} $ & $ 9.9 \pm 0.08 $ \\
\hline
   $K^{\pm} $ & $ 2.42 \pm 0.13 $ \\
\hline
   $K^{0} $ & $ 2.12 \pm 0.06 $ \\
\hline
   $\eta $ & $ 0.73 \pm 0.07 $ \\
\hline
   $\rho^{0}(770) $ & $ 1.4 \pm 0.1 $ \\
\hline
   $K^{*\pm}(892) $ & $ 0.78 \pm 0.08 $ \\
\hline
   $K^{*0}(892) $ & $ 0.77 \pm 0.09 $ \\
\hline
\end{tabular}
\parbox{7.cm}{\caption{Production rates of light mesons
 in hadronic $Z^0$ events as measured at LEP
 (Table 1, taken from \protect\cite{tun}). We see 
 that most of the BE effect can be expected from correlations between
   pions, eventually kaons (the production rates for other bosonic
 species are rather low).}
\label{tabpr}}
\end{minipage}%
\begin{minipage}[t]{8.cm}
\begin{tabular}{|c|c|}
\hline
 Origin of $\pi^{+}$ & Fraction [\%] \\
 in $Z^{0}$ decay   &  (JETSET 7.4) \\
\hline
  direct  &  16 \\
 ( string fragmentation) & \\
\hline
  decay of short-lived resonances  & 62 \\
      $\Gamma > 6.7$ MeV        &    \\
  ($\rho,\omega,K^{*},\Delta,\ldots$) & \\
\hline
  decay of long-lived resonances  & 22 \\
      $\Gamma < 6.7$ MeV        &    \\
\hline
\end{tabular}
\parbox{8.cm}{\caption{ The origin of charged pions
   in hadronic $Z^0$ decay. The table shows
   how many of charged pions come directly from
   string fragmentation and from decay of resonances ( the division between
   short and long-lived resonances is arbitrary, here it corresponds to
 a life-time of about 30 fm/c).}
\label{tabpr2}}
\end{minipage}   
\end{table}

   Among all bosons produced in the event, mainly direct hadrons 
 (products of string fragmentation) and decay products of shortly living
 resonances are susceptible to be influenced by BE correlations.
  We have included BE interference for the following bosons: $ \pi, K, \rho$
 and  $\omega $. Every prompt boson of one of these types goes through
 a local reweighting procedure at the moment of its generation, e.g.
 at the moment of string fragmentation or at the moment of the decay
 of the mother resonance.
 The string fragmentation cycle itself is not disturbed; all
 direct hadrons coming from a single string are reweighted together,
 which means that the fragmentation of each string is repeated until
 the correlation function -- the product of sums of interference
 terms for all identical bosons -- passes weighting criterium.

   The decay of a short-lived resonance is affected
 by local weighting if -- among its decay products -- there are identical bosons
 or bosons of the same type as those already generated. (We call the weighting
 ``local'' to stress the fact that -- contrary to the global weighting --
 we split the total correlation function ( the global weight ) into a set
 of separate ``local'' weights.)
 The energy and momentum of the mother resonance is preserved,
 as well as the decay channel it started to decay into, while its life-time is
 allowed to vary. The weighting is used to find,
 in the available phase space and according to the correlation function,
 the configuration where
 daughter bosons are close to bosons already existing. We would like 
 to point out the fact that there is no double counting of
 interference terms, and that the order of generation is actually irrelevant,
 since the individual terms in the total correlation function are
 Lorentz invariant.

 An option is included in the MC program which allows  the decay 
 products of a resonance to be treated as if they were
 direct hadrons. Especially in the case of $\rho$ mesons, the resonances
 decay so quickly that their decay should be actually treated as  part
 of string fragmentation. In practice however this option is of little use: 
 the more direct bosons we have, the less effective the weighting is,
 and in addition the multiplicity of direct hadrons runs out of control.
  
  The whole procedure is rather intuitive --  the probability
 of having close bosons is enhanced step by step until the complete
 final state is
 generated, while most of the standard JETSET features are preserved. 
     The method is obviously more effective than the global weighting 
 \cite{markus}, however
  the overall normalization scale being lost,  we don't know a priori
 how many close pairs and triplets are needed to reproduce the experimental
 data. As we will see in the next section, the method of ``forced'' generation
 of close bosonic pairs, a priori expected to give a somewhat exaggerated
 BE correlations, seems to agree rather well with experimental observations.

\begin{figure}[bth]
\begin{center} 
\mbox{\epsfig{file=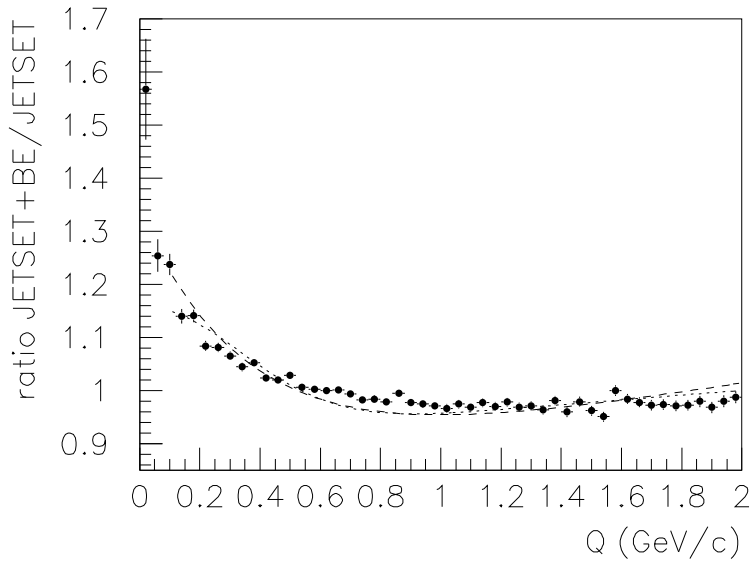,height=10.cm}}
\end{center} 
\parbox{15.cm}{\caption{\protect\small \sl
 Simulated two particle correlation function for
 like-sign pairs compared to fits of the DELPHI 
 data (simulated sample of $10^5$ events).}  
\label{fig:qz}}

\vspace{0.2cm}

\parbox{15.cm}{ \sl
 Gaussian fit to DELPHI data (dotted line):  
 $ 0.91 \ (1.+ 0.05 \ Q) (1.+ 0.27 \ \exp{\{-(2.16 \ Q)^2\}} ) $  \\
 Exponential fit to DELPHI data (dashed line):
 $  0.83 \ ( 1. + 0.11 \ Q ) ( 1. + 0.61 \ \exp{\{- 2.82 \ Q\}} ) $}
\end{figure}

\section{Results of simulation and 
 comparison with experimental measurements } 

   Fig.~\ref{fig:qz} shows the two-particle correlation
 function for like-sign pairs of particles from $Z^0$ decay
 ($\mathrm{E}_{CMS}=91.22$ GeV),
 obtained with our simulated data. The variable $ Q = \sqrt{-(q_1-q_2)^2} $
 is the momentum transfer between two particles with momenta $q_1,q_2$.
 Only particles with  momentum above 0.2 GeV/{\it c} were taken into
 account, and -- similarly to the experiment -- the decay products
 of $K^0$ and $\Lambda$ were removed.
 For comparison,  fits to the DELPHI data with an exponential and with a 
 gaussian parameterization are plotted as well \cite{BE-DELPHI}.

  The simulation reproduces the enhancement of the two-particle
 correlation function rather well. There is a small discrepancy: 
 a small linear rise of the correlation function
 with $Q$ is observed in the data, but not in the simulation. This effect is
 most probably due to a residual difference between the reference sample
 for the data (which is a sample of tracks mixed from different events)
 and for simulation (represented by the JETSET simulation without the BE
 correlation). 

\begin{figure}[bth]
\mbox{\epsfig{file=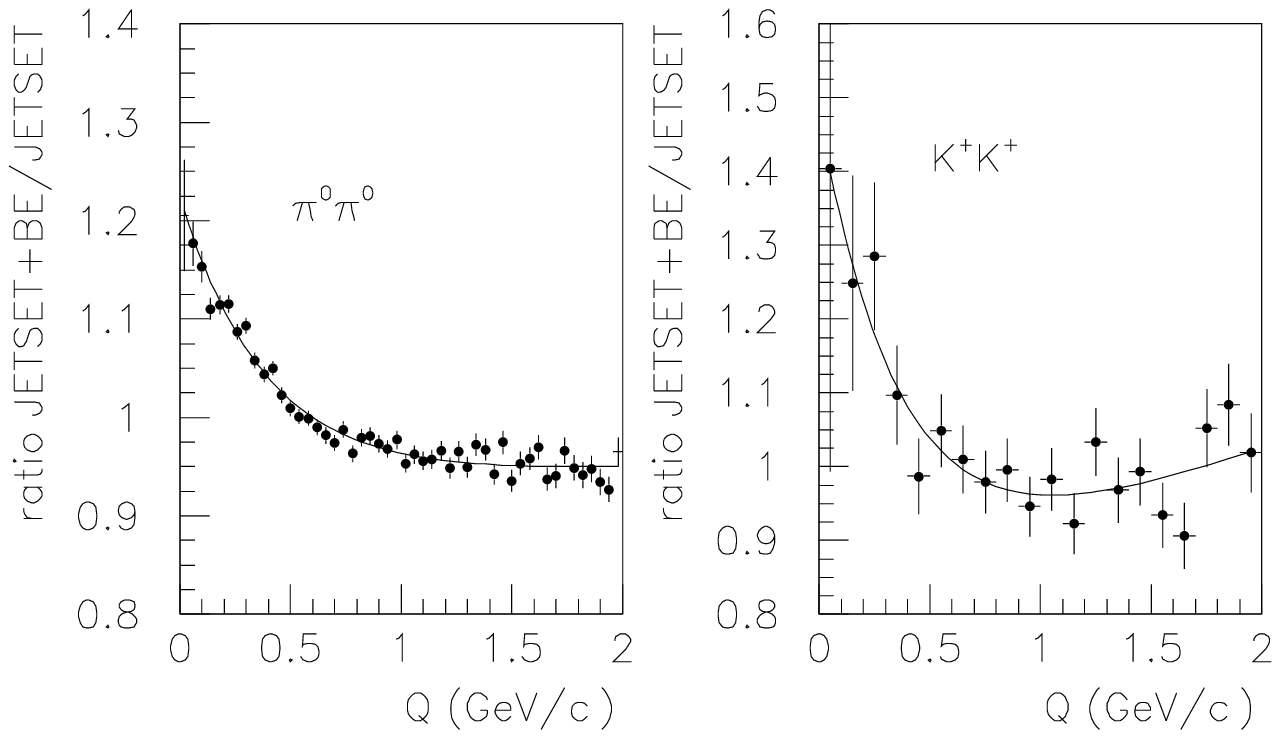,height=10.cm,width=15.cm}}
\parbox[t]{6.cm}{\caption{\protect\small \sl
 Simulated 2-particle correlation 
 function for pairs of neutral pions. 
 Fit: $ 0.94 [1.+ 0.3 \ exp (-2.6 Q) ] $ } 
 \label{fig:qkpi1}}
\hspace{1.cm}
\parbox[t]{6.cm}{\caption{\protect\small \sl
 Simulated 2-particle correlation 
 function for pairs of charged kaons. 
 Fit: $ 0.8 (1.+ 0.1 Q) [1.+ \ exp (-2.8 Q)] $ } 
\label{fig:qkpi2}}
\end{figure}

    The simulated two particle correlation functions for neutral pions
 and for charged kaons are shown in Fig.~\ref{fig:qkpi1},~\ref{fig:qkpi2};
 both were fitted
 with exponential parameterizations.

\begin{figure}[h]
\mbox{\epsfig{file=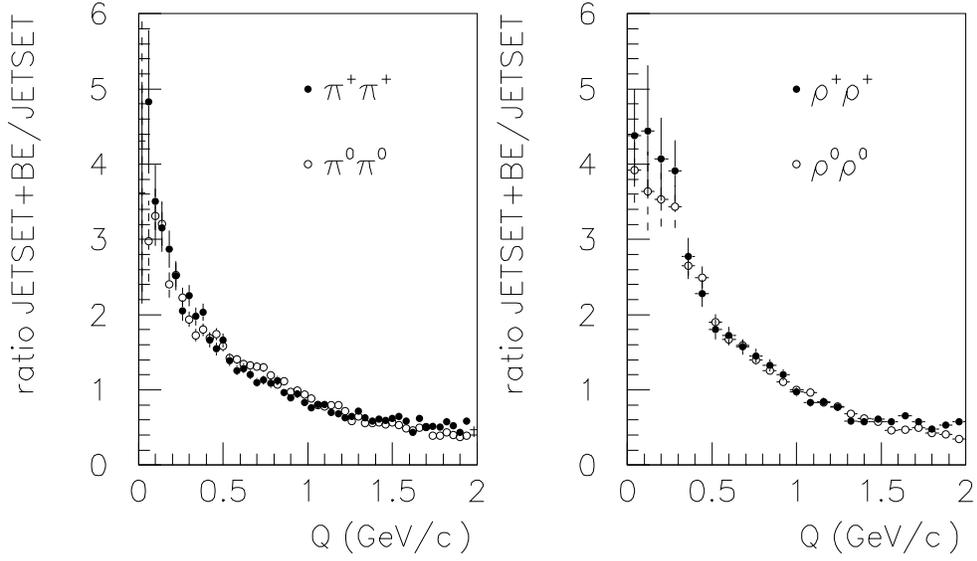,height=10.cm,width=15.cm}}
\parbox[t]{15.cm} {\caption{\protect\small \sl
 Simulated 2-particle correlation 
 function for pairs of direct bosons. }
\label{fig:direct}} 
\end{figure}

\begin{figure}[h]
\mbox{\epsfig{file=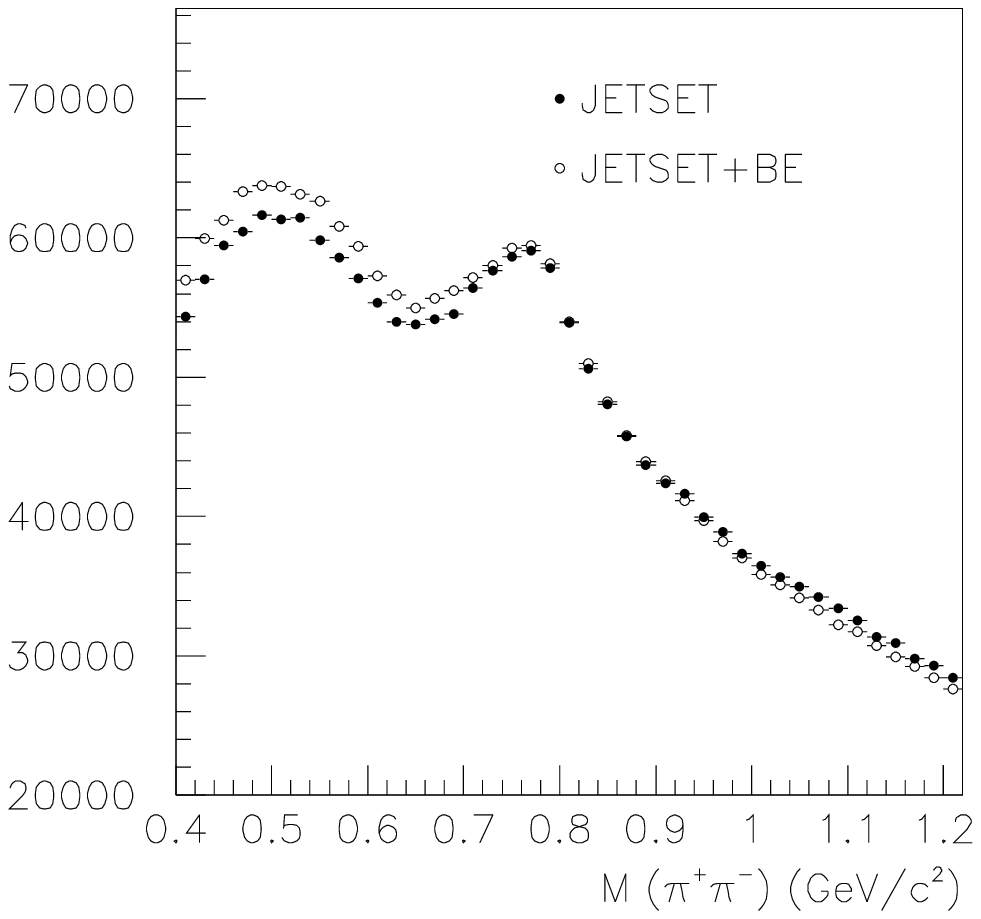,height=8.cm,width=8.cm}
      \epsfig{file=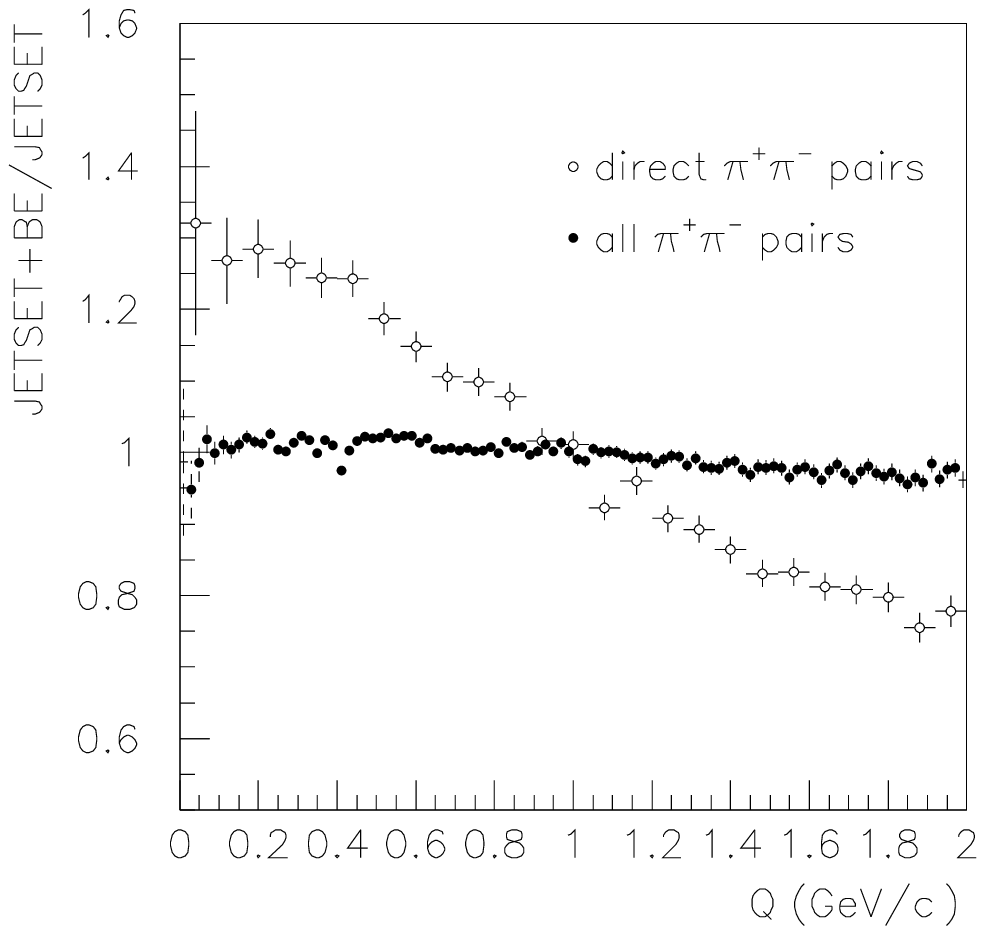,height=8.cm,width=8.cm}}
\parbox[t]{7.cm} {\caption{\protect\small \sl
 Mass distribution of prompt  
$\pi^+\pi^-$ pairs. }
\label{fig:mpp}} 
\hspace{1.cm}
\parbox[t]{7.cm} {\caption{\protect\small \sl
 Simulated 2-particle correlation function for $\pi^+ \pi^-$ pairs.}
\label{fig:mixed}} 
\end{figure}

\begin{figure}[h]
\mbox{\epsfig{file=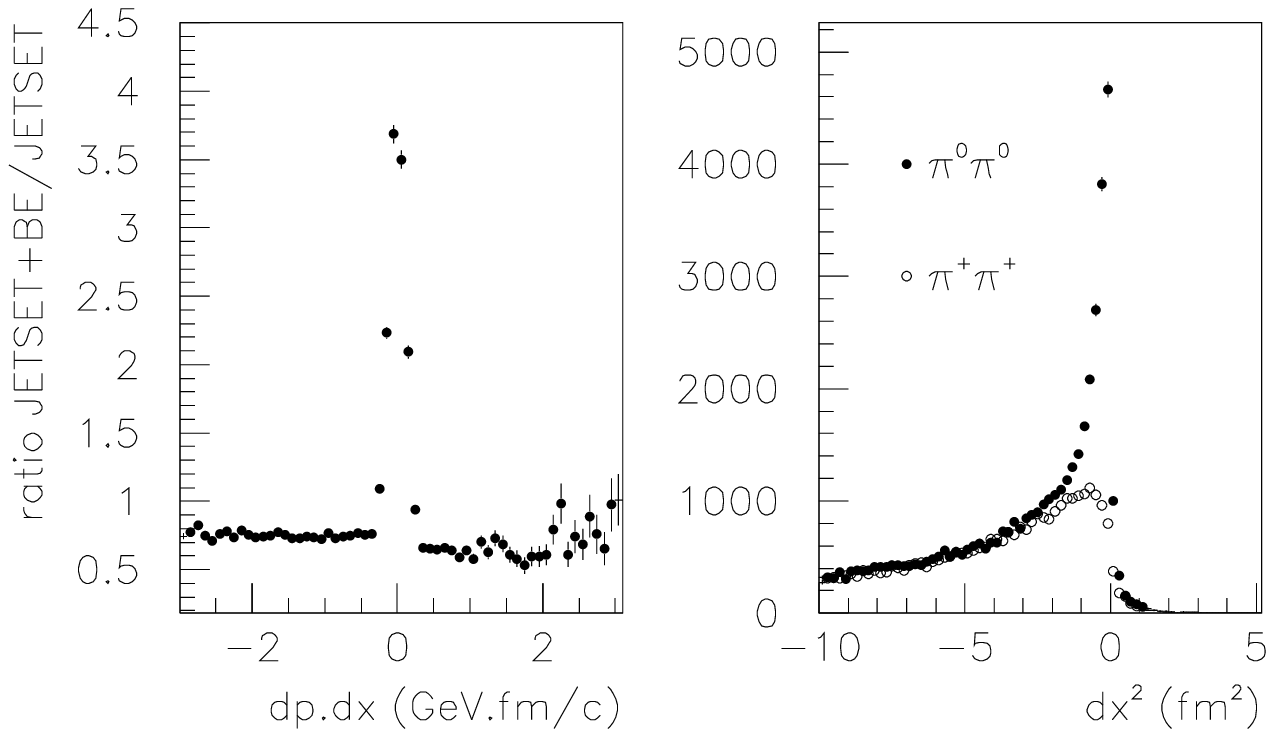,height=8.5cm,width=15.cm}}
\parbox[t]{7.cm} {\caption{\protect\small \sl
 The $dp\cdot dx$ distribution of
 prompt equally charged pions modified by BE correlations. } 
\label{fig:dpdx}}  \hspace{0.5cm}
\parbox[t]{7.5cm} {\caption{\protect\small \sl
  Distribution of the squared space-time distance  between
 pairs of direct pions (JETSET without BE correlations).} 
\label{fig:dis}} 
\end{figure}

\begin{figure}[h]
\mbox{\epsfig{file=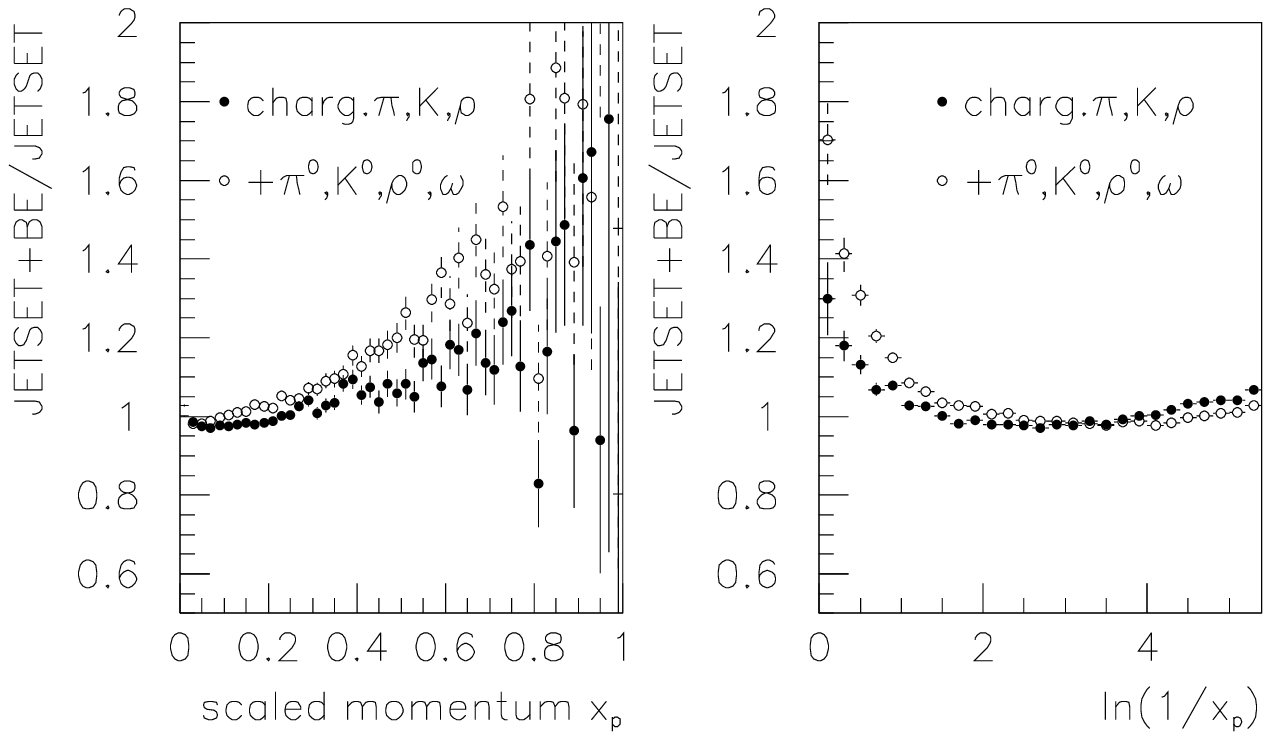,height=8.5cm,width=15.cm}}
\parbox[t]{15.cm} {\caption{\protect\small \sl
  Scaled momentum distribution of charged particles and its logarithm,
 modified by BE correlations between 
 charged and between charged+neutral prompt bosons.
 } 
\label{fig:xp}}
\end{figure}

  The enhancement in the production of close pairs of direct bosons is strong
 (see Fig.~\ref{fig:direct}), but most of the effect observed in final
 hadronic states is due to the correlation between pions from resonance
 decays .
 This leads us to another observable feature of BE interference,
 namely the possible distortion of the resonance spectrum (observed at
 LEP for the $\rho^0$ \cite{rho}).

   The mass spectrum of direct resonances is in principle allowed
 to vary in our approach. Even so, we don't observe any significant 
 change in the $\rho^0$ spectrum itself, contrary to \cite{markus}.
 What we do see,
 however, is a non-negligible modification of the ``background''
 $\pi^+ \pi^- $  
 spectrum, clearly influenced by BE correlations between
 identical bosons, and which would lead to a lower fitted value of the $\rho^0$
 mass if not taken into account (Fig.~\ref{fig:mpp}). 
   The two-particle correlation functions for direct and for all
 $\pi^+ \pi^- $ pairs in the
 final state are shown in Fig.~\ref{fig:mixed}.  
 
  Although we have strongly influenced the distribution of identical bosons
 in the configuration space (Fig.~\ref{fig:dpdx}),
  the changes in  event shape variables are not very dramatic. Part of
 them are directly related to the change of the charged multiplicity:
 when weighting the products of the string fragmentation we don't
 fix the multiplicity of direct bosons and therefore we partly
 loose the control over the multiplicity of the final state.
 The total charged multiplicity
 decreases by $ 5\% $ if correlations are included for all bosons mentioned
 above; it increases by $ 2\% $ if only charged bosons are taken into
 account (because identical neutral bosons can be produced
 at closer space-time distance, and are therefore  more easily correlated
 than equally charged bosons, see Fig.~\ref{fig:dis}).
 Fig.~\ref{fig:xp} shows  the  behaviour
 of the scaled momentum distribution of charged final particles, both
 in the case where only charged bosons are correlated and in the case
 when neutral bosons are correlated as well.
 The distribution is enhanced at both  ends of the spectrum -- 
 a feature supported by the data \cite{tun}. It would probably be worthwhile
 to retune the JETSET parameters in order to see how much of this effect 
 remains when the total charged multiplicity is adjusted.
 
\clearpage

\section{BE effect and factorial moments}

An alternative tool to ordinary correlation functions  
in studies of particle correlation is represented by factorial moments.
Originally this notion was introduced by Bia{\l}as and Peschanski in 1986
\cite {BiaPe} in connection with intermittency. Roughly speaking, the
 underlying question 
was whether the fluctuations of local density of some quantity 
(like rapidity, azimuthal angle, transverse momentum) are of purely 
statistical nature, or have some non-trivial origin. The factorial moments
 have been
shown to provide the suitable method for addressing these problems. 
     
An $i-th$ factorial moment can be defined as 

\begin{equation}
F_{i}=\frac{1}{N_{events}}\sum_{events}
\frac{\sum_{k=1}^{n_{bins}} \left\{n_{k}(n_{k}-1)
\cdots (n_{k}-i+1)\right\}/n_{bins}}
{(\langle n\rangle /n_{bins})^{i}}
\label{genfacmom}
\end{equation}
where $\langle n \rangle$ is the average number of particles in the full
phase space region accepted, $b_{bins}$ denotes the number of bins in
this region, which is given by $(2^b)^d,\;b=0,1,2...$ ($d$ is the  
dimension of the phase space region considered) and $n_k$ is the
multiplicity in $k$-th bin. In what follows we consider factorial
moments in two and three phase-space dimensions, in the conventional
variables $(y,\varphi)$ and $(y,\varphi,\tilde{p_t})$ ($y$ denotes rapidity, $\varphi$ 
is the azimuthal
angle and $\tilde{p_t}$ is connected to the transverse momentum -- 
 it is defined as in \cite{ptilda}
\begin{equation}
\tilde{p_t}=\frac{\int_{0}^{p_t}P(p_t)dp_t}{\int_{0}^{p_t^{max}}P(p_t)dp_t}
\label{ptilda}
\end{equation}
where $P(p_t)$ is the probability distribution of $p_t$ in the interval
$(0,p_t^{max})$, $p_t^{max}$ being some suitably chosen upper limit. The purpose of 
treating $p_t$ this way (and in principle any other quantity of highly non-uniform 
density distribution - $p_t$ itself is steeply falling)  
is to make the overall distribution of a quantity studied more uniform, 
which is a necessary condition for this type of analysis \cite{ptilda}.

The method originally proposed in \cite{BiaPe} consists in measuring the dependence 
of factorial moments defined in Eq.\ref{genfacmom} as a function 
of ``resolution''
in phase space, i.e. of $b$ in our notation. The statement is that while purely 
statistical fluctuations in density lead to constant behaviour of $F_{i}$ with respect
to $b$, the presence of non-trivial correlations is signalized by its rise. The content
of original concept of ``intermittency'' was even stronger -- that the rise in double-log
scale should be linear, i.e. $ \ log(F_{i}(b)) \propto \phi_{i}b $, where the 
``intermittency
index'' $\phi_{i}$ had been claimed to be connected with the fractal character of hadron
or underlying parton shower and various dynamical models of fragmentation fulfilling
these conditions had been formulated (see \cite{DeWolf} and references therein).

 \begin{figure}[bth]
\epsfig{file=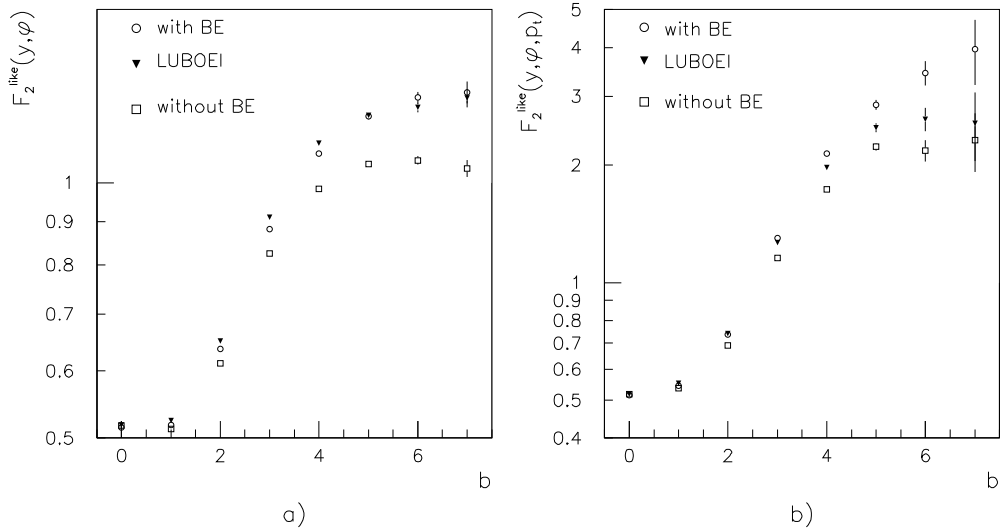,width=15.cm}
\parbox{15.cm}{\caption{Factorial moments $F_2^{like}$, for two
(Fig.\ref{fig:fm}a) and three (Fig.\ref{fig:fm}b) phase space variables
 (see text), calculated by
 JETSET for hadronic decay of $Z^0$ at the centre of mass energy $91.22$ GeV,
 with and without the
 BE correlations. For comparison, the result of the simulation
 with LUBOEI (with parameters $PARJ(92)=0.35$, $ PARJ(93)=0.42 \ 
 \mathrm{GeV} $
 in the gaussian
 parameterization) is shown as well.}
\label{fig:fm}}
\end{figure}

At present the outlook somewhat changed and it is generally accepted that there
is no proper ``intermittency'' in the above sense and BE correlations are the only cause
of the short range correlation (e.g. \cite{BiaVietri}, \cite{AmiDEL} ). Simultaneously,
the quantities studied shifted from factorial moments to other ones, mostly two-
or more-particle correlation functions in various phase-space variables. This however
does not change the basic fact that the presence of any correlations should be observable
in terms of factorial moments as well -- it should lead to their rising as phase space
variables bins decrease (though, with the original concept of intermittency all but 
abandoned, there is no deeper interpretation of its slope and even no deeper reasons why  
it should be linear in the double-log scale at all).
There are even some advantages in comparison with
``ordinary'' correlation functions, like reasonably straightforward (from the technical
point of view) construction of two- or three-dimensional (in phase-space variables) 
quantities and no need to construct the ``uncorrelated'' ensemble for
normalization. 

We will study, with the help of factorial moments method, the two-particle correlations 
of particles generated with the BE correlations switched on and off, 
respectively,
in the JETSET generator. We define the like-charge second
 factorial moments as 

\begin{equation}
F_{2}^{like}=\frac{1}{N_{events}}\sum_{events}
\left(\frac{\sum_{k=1}^{n_{bins}} \left\{n_{k}^{+}(n_{k}^{+}-1)
\right\}/n_{bins}}
{(\langle n^{+}\rangle /n_{bins})^{2}} + 
\frac{\sum_{k=1}^{n_{bins}} \left\{n_{k}^{-}(n_{k}^{-}-1)
\right\}/n_{bins}}
{(\langle n^{-}\rangle /n_{bins})^{2}}\right) 
\label{Fstejny}
\end{equation}

where $n_{k}^{+}$ and $n_{k}^{-}$ denotes the number of positive and negative particles,
respectively, in the $k-th$ bin, while $\langle n^{+}\rangle$, $\langle n^{-}\rangle$
are the average numbers of positive and negative particles in the full phase space
region. In calculating factorial moments, we restrict ourselves to the rapidity interval 
$-3.2 \leq y \leq 3.2$ and we take $p_t^{max}$ (see Eq.\ref{ptilda})
 to be 2 GeV/{\it c}.
The phase space
variables are expressed with respect to the thrust axis of each event.

The results are shown in Fig.\ref{fig:fm}. We calculated,
 with the help of the JETSET generator,  
the factorial moments for like-charge particles 
(Eq.\ref{Fstejny}) produced in the decay of $Z^0$. 
On Fig.\ref{fig:fm}a we can see the behaviour of the two-dimensional
 moments $F_{2}^{like}(y,\phi;b)$
as a function of $b$ (see Eq.\ref{genfacmom}),
Fig.\ref{fig:fm}b  shows the same for 
three-dimensional moments $F_{2}^{like}(y,\phi,\tilde{p_t};b)$. The three sets
 of points on each plot have been calculated from data generated
 without any BE correlations, with the BE correlations
 included according to the original JETSET option (subroutine LUBOEI), and with 
the BE correlations implemented  as described in the present paper,
 respectively. We can see that the treatment of the  BE
correlations based on the space-time picture of the production process
 leads to the right effect: the factorial moments rise.
 It should not be surprising that the effect is more pronounced
 for the three-dimensional moments than for the two-dimensional ones,
 as projection of the correlation effect onto the lower dimension subspace
 can ``dilute'' the effect
 and lead to the flattening of the behaviour of the moments \cite{DeWolf}.

  In the simulation with LUBOEI, the gaussian parameterization was used
 with parameters  $ PARJ(92) = 0.35$,
 $PARJ(93) = 0.42 \ \mathrm{GeV}$, which is in agreement with data \cite{BE-DELPHI}.
  There is practically no difference
 between the two methods of the treatment of the BE correlations
 in the two-dimensional case. However,
  the three-dimensional factorial moments calculated 
 with the original JETSET BE recipe (LUBOEI) seem to behave similarly to those
 calculated with no BE correlations, e.g. they reach a plateau. A further rise
 of the three-dimensional factorial moments in data generated with LUBOEI
 could be achieved by setting its parameters to higher values, but this
 would imply much stronger enhancement of the 2-particle correlation 
 function than observed in the experimental data.

\section{Does the BE effect influence the measurement of the W mass
 at LEP2 ? }

   The study of hadronic WW events certainly adds a new dimension
 to the problematics of the BE interference. Now we have to deal with
 -- at least -- two strings . In fact, we had a multiple string configuration
 in $Z^0$ decays as well -- as the result of a gluon splitting -- but 
 since our weighting algorithm is based
 on the calculation of the absolute coordinates of a hadron position, it can
 -- technically -- handle such a configuration without difficulties,
 and we actually didn't ask how should the BE interference look like for
 bosons from different strings.
   Nevertheless, in the study of the systematic error on the W mass, 
 this question requires a detailed  discussion.

 \begin{figure}[bth]
\mbox{\epsfig{file=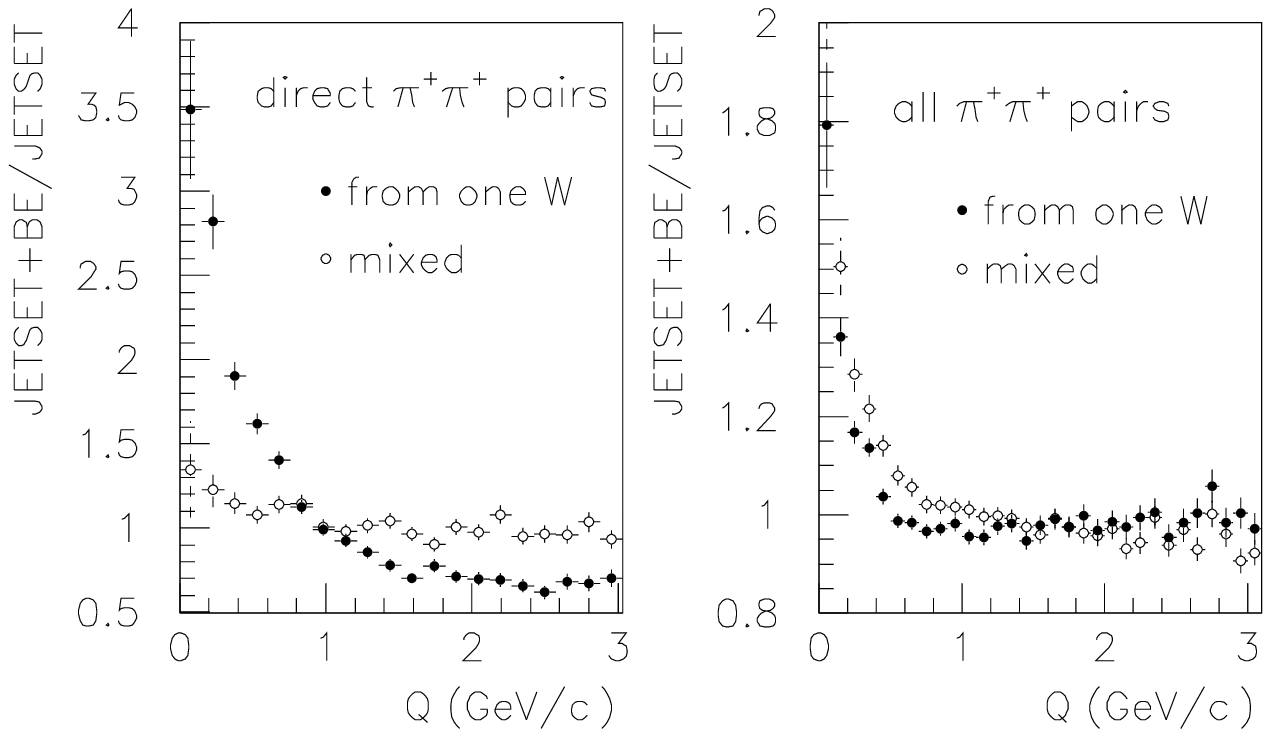,height=8 cm,width=15cm}}
\parbox{7.cm}{\caption{\protect\small \sl
 The simulated two-particle correlation function for pairs 
 of direct equally charged pions
 coming from the same W and those of a mixed origin.
 WW hadronic events generated at 172 GeV.} 
\label{fig:ww1}}
 \hspace{1.cm}
\parbox{7.cm}{\caption{\protect\small \sl
 The simulated two-particle correlation function for pairs 
 of all equally charged pions
 coming from the same W and those of a mixed origin.
 WW hadronic events generated at 172 GeV.} 
\label{fig:ww2}}
\end{figure}

   To avoid confusion, we start with the discussion of the relationship between
 the BE effect and colour reconnection (often they are put together and called
 interconnection effects). Colour reconnection is the term used for the
 interaction of strings which changes the string configuration
 ('reconnects' them),  and therefore implies momentum/energy transfer
 between the original strings. On the other hand, while deriving the correlation
 function for the BE effect, we didn't account for any explicit interaction
 term between
 different strings. In fact, we derived it {\em only} for a single
 string. While Eq.2 can be -- at least formally -- applied to bosons
 coming from different strings, this formula doesn't contain the
 exponential suppression and, when actually used, does not produce any
 observable effect in the simulated data. Therefore we consider
 the BE interference as preserving the total string momentum
 and every direct string-string interaction with
 momentum transfer will be considered as belonging to colour reconnection. 
 The interplay of the BE effect and colour reconnection can be investigated 
 with the help of existing phenomenological models for simulation of
 colour reconnection (those based on JETSET fragmentation can be combined
 with simulation of BE interference without difficulties
 \footnote{The influence of colour 
 reconnection on W mass measurement was investigated in \cite{CR,me,JZ}. }).

   In agreement with the classification introduced above, the mass of the string
 is preserved during hadronization. Which are then the remaining possibilities
 to see the W mass spectrum modified? One of them is purely
 experimental and concerns only fully hadronic WW events:
 since we are not able to separate completely the two
 hadronic systems (one belonging to the $W^+$, the other to the $W^-$), there
 is always a fraction of misassigned particles resulting in a smearing 
 of the measured W mass spectrum. The Monte-Carlo simulation can be used 
 to correct for this effect. The Bose-Einstein effect, however,
 with its tendency to produce boson pairs with similar momenta,
 can change the fraction of misassigned particles; if this effect would be 
 missing in the simulation, we would obtain a wrong estimate of the 
 correction to apply to the observed mass. 

    Another possibility to get a distorted spectrum is more fundamental,
 if we admit that the primary process itself (the production of WW pairs)
 may be influenced by the interference terms added to the hadronization part.
 It seems however unlikely to be so; after all, the whole simulation of 
 the hadronization makes use of the so called factorization theorem: the 
 amplitude of string
 fragmentation (Eq.~\ref{eq:amp}) doesn't appear in the total event
 weight nor is the hard
  process or parton configuration rejected because of fragmentation.
    Still, we don't see really strong arguments 
 why the hard process should not be influenced, and therefore we made a
 check of what happens with the W spectrum if we use our weights for direct
 hadrons as the global event weights for the sample of semileptonic WW events.
 A sample of 500,000 events was generated with PYTHIA/JETSET including our BE
 simulation. The reweighted spectrum of the hadronic W mass was 
 compared to the
 generated one (both were fitted with a Breit-Wigner distribution times
 a phase space factor). The result is shown in Table \ref{wmass}
 (method I). 
  
\vspace{0.5cm}
 
\begin{table}[tbh]
\begin{tabular}{|c|c|}
\hline
  Method ($E_{CMS}$=172 GeV) & Shift of fitted W mass [MeV]\\
\hline
I:  weight for direct bosons &      \\
  used as global event weight &  $ -10 \pm 12 $     \\
  (in semileptonic WW events) &       \\
\hline
II:   BE interference included only within a string  & $ +11 \pm 11 $     \\
  (unweighted sample,hadronic WW events) &  \\
\hline
III:   BE interference among strings as well  & $ +12 \pm 11 $       \\
  (unweighted sample,hadronic WW events) &  \\
\hline
\end{tabular}
\parbox{15.cm}{\caption{The shift of the fitted W mass due to the BE effect
 in various scenarios (see text).}
\label{wmass}}
\end{table}

\vspace{0.5cm}
 
  Since this is the method which is closest to the use of global weights
 in \cite{RM},
 we have also checked the effect of this weighting on the values of $R_b,R_c$
 in  $Z^0$ decays. We observed a (statistically insignificant)
 difference of the order of a few per-cent
 ($+5\pm2\%$ for $R_c$, $+1\pm2\%$ for $R_b$). 

 We don't feel that 
 doing the same exercise with fully hadronic events is useful -- the
 interference across different strings is really ill defined for such a study.
  It can be nevertheless used to study the experimental
 problem of wrongly assigned
 particles, because it mimics rather well the situation when
 (for some reason) independent strings produce ``mixed'' 
 pairs of bosons of similar momenta (as if Eq.\ref{eq:sup} would be valid
 for all bosons in the event). In fact, two studies were made:
 one with BE correlations allowed only inside a single string, the other
 with correlations of bosons coming from different strings included as well
 (methods II and III in Table~\ref{wmass}).
 In each event, the mean W mass was
 reconstructed (only clear four jet events were used, i.e. with a 
 minimal energy per jet of 20 GeV and an angular separation between jets larger
 than 0.5 rad), then the mass distribution was fitted and compared
 to the reference sample 
 (standard PYTHIA/JETSET without BE correlations). The results are also
 shown in Table \ref{wmass}.

    For illustration, Fig.~\ref{fig:ww1},~\ref{fig:ww2}
 show the two-particle correlation
 function for pairs of pions from decays of the same W and for those of
 'mixed' origin.
  We remind once more that while the calculation of weights for mixed pairs
 is technically straightforward in our approach (which is based 
 on the evaluation of hadron's production vertex), their use is not warranted
 by the QM arguments as for bosons coming from a single string.
 (Indeed, the very first results of measurements of BE correlations 
 in WW events at LEP \cite{wwmes} suggest that the interference between
 strings/W's is strongly suppressed.)

  The results of our studies do not signal any special danger for the W mass 
 measurement; we don't see
 how the BE effect can shift the W mass by 50 or even 100 MeV as suggested
 in \cite{YR}, 
 even when we take the interference between the different
 strings to be as strong as the interference inside a single string. 
 The uncertainty quoted in Table \ref{wmass} is based on the statistical error
 of the fit of the W mass distribution and could be decreased just by
 increasing the simulated sample. However, taking into account other related
 uncertainties (the study is done at the generator level,
 the reconstruction method we use does not
 necessarily correspond to the one actually used in the experiment, the shape
 of reconstructed W mass distribution is not a simple Breit-Wigner 
 distribution convoluted
 with phase-space factor and so on), we think the quoted error is a realistic
 one. 
 
   To make the picture more complete, we would like to investigate a little
 bit more the space-time picture of hadronization. The very general
 argument why there should be some interference between the two W's
 says that because the W's decay close to each other, the strings overlap and
 are very likely to have some sort of interaction. Let us take the example
 of an ordinary hadronic WW event at 172 GeV:
  the W's decayed at a distance of 0.05 fm, their decays were 
 followed by parton showering and there are two or more strings around
 evolving towards fragmentation (the mean life-time of a string is about 
 1.5 fm/{\it c}).
 The two hadronic systems are separating
 (the mean velocity for W's is around 0.4 {\it c}), the decay planes of both
 W's being different. It is therefore not so evident that strings have to be
 in contact.
 In fact, the colour reconnection study \cite{me} shows that in nearly $40\%$ 
 of all events, the overlap of strings is negligible.

\begin{figure}[bth]
\mbox{\epsfig{file=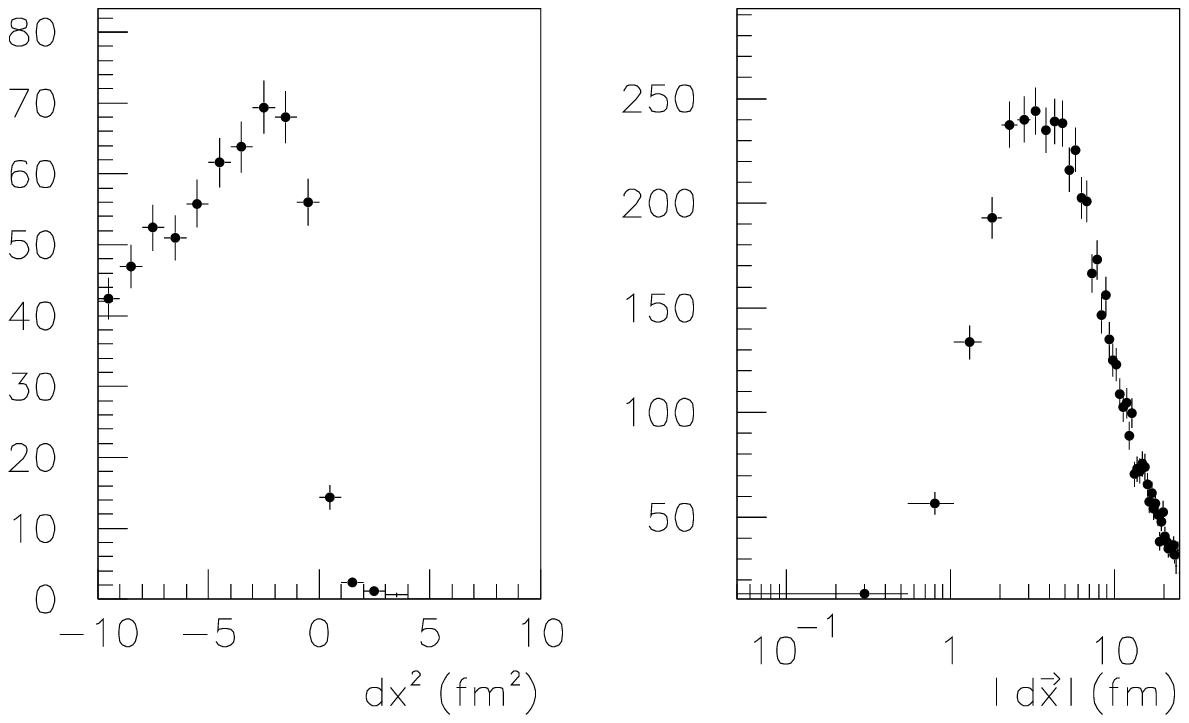,height=8 cm,width=15cm}}
\parbox{7.cm}{\caption{\protect\small \sl
 The space-time distance squared of production vertices of direct equally
 charged pions for pairs of mixed origin. WW hadronic events at 172 GeV.
} 
\label{fig:wwdx2}}
 \hspace{1.cm}
\parbox{7.cm}{\caption{\protect\small \sl
 The space distance  of production vertices of direct equally
 charged pions for pairs of mixed origin. WW hadronic events at 172 GeV.
} 
\label{fig:wwdx}}
\end{figure}

 Since we believe that the origins of the BE effect lie
 somewhere in the fragmentation, we are interested how often strings
 do overlap while fragmenting.  Fig.~\ref{fig:wwdx2} shows the square of
 the space-time interval between production vertices  of equally charged
 direct pions for mixed pairs (one pion coming from the $W^+$, the other
 from the $W^-$), while Fig.~\ref{fig:wwdx} shows the distance in space coordinates
 only. The production vertices are causally
 disconnected and the mean distance between them exceeds the typical
 transverse size of a string (about 1 fm): there is no evidence
 of a sizeable overlapping of strings.

\clearpage 

\section{BE effect as flavour correlation in string fragmentation } 

    The simulation model we have presented
 certainly does capture some important features of the BE effect and may
 be of some use for practical purposes; it is not limited by the topology
 of events (e.g. multiple jets), has a relatively short execution
 time and can be developed further.  Still, it is not the best solution
 for the problem of BE simulation, and this for several reasons discussed
 hereafter.
 
    First, a rather important amount of computing time is spent on
 generating and rejecting  events which do not contain any bosons
 with similar momenta. The weighting procedure makes us {\em  to wait}
 for the accidental generation of events which we know  -- more or less --
 how they will look like. Second, we are to some extent loosing control
 over some important parameters, like the multiplicity of the final state.
 We can in principle react by retuning the parameters of the model, but we
 risk to be confronted with this kind of problem again and again; in short, 
 the approach is {\em inconsistent} with the philosophy of the fragmentation
 model.

    We would like to devote this section to a discussion about a potential
 new approach to BE simulation -- the direct implementation
 into the fragmentation scheme.
 Not that we have the complete solution on hand, but we are
 convinced that such a simulation is feasible. It would require some changes
 in the fragmentation model but would pay off in the long term.

\begin{figure}[bth]
\mbox{\epsfig{file=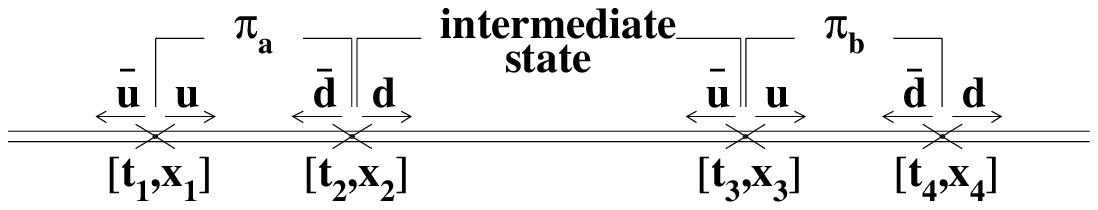,height=5cm,width=15cm}}
\parbox{15.cm}{\caption{\protect\small \sl
} 
\label{fig:flav}}
\end{figure}

   To show what we have in mind, let's take once more the case of
 a simple $q\tilde{q}$ string in its rest frame.  It will fragment into a set
 of hadrons, among them two identical bosons $a,b$
 ( let's say charged pions, for definiteness), as in Fig.~\ref{fig:flav}.

\vspace{-2.cm}

  We will calculate the $dp\cdot dx$ term for the pair (a,b)
 using Eq.\ref{eq:f1}
 and \ref{eq:f2} :
\begin{eqnarray}      
   (p_a - p_b)\cdot (x_a - x_b) & = & (E_a - E_b)(t_a - t_b) - 
                       (\vec{p}_a - \vec{p}_b)(\vec{x}_a - \vec{x}_b) 
                               \nonumber \\
                                & = & 0.5 (x_2 -x_1 -x_4 +x_3) 
          ( t_1 +t_2 - t_3 - t_4)  \nonumber \\
                                &   & - 0.5 (t_2 -t_1 -t_4 +t_3)
          ( x_1 +x_2 - x_3 - x_4) \nonumber \\
                                & = & \cdots \nonumber \\
                                & = & (x_4 -x_2) (t_3 -t_1) - 
            (x_3 -x_1) (t_4 -t_2) \nonumber \\
                                & = & E_{d}  p_{u_{l}}
                     - E_{u} p_{d_{l}} 
\label{eq:flc}
\end{eqnarray}
   and we have a relationship between the Bose-Einstein effect and
 the flavour correlation in string fragmentation; the correlation depends
 on the energy-momentum of string pieces for which flavour is compensated
 (i.e. the end-point quarks are of the same flavour). 
 
      Now we can switch to the light-cone metrics in which the Lund
 fragmentation model is formulated: the Lorentz invariant variable
 $z^+ (z^-)$ determines
 the fraction of energy-momentum of the end-point (massless) quark(antiquark)
 which the hadron takes away:
\begin{eqnarray}
  E_h & = & ( z^+ + z^- ) E_{q0} \nonumber \\
  p_h & = & ( z^+ - z^- ) p_{q0} 
\label{eq:fez}
\end{eqnarray}
 In the Lund fragmentation model, the hadron is defined in 3 steps:
\begin{enumerate}
\item  the flavour of the next string breaking is chosen, as well as the
 hadron mass {\em m};
\item  the transverse momentum of the new quark/antiquark pair is generated
 (according to a gaussian distribution), defining the total transverse
 momentum $p_t$ of the hadron;
\item  $ z^+ $ ( or $ z^- $) is generated according to the Lund symmetric
 fragmentation function ; the momentum of the hadron is thus fully determined
 (the remaining $ z^- $($ z^+ $) is calculated from the relation
\begin{equation} 
   z^+ z^- M^2_{0} = m^2_t = m^2 + p^2_t 
\end{equation}
 where $M_0 = 2 E_0$ stands for the mass of the string). 
\end{enumerate} 

 With the help of Eqs.\ref{eq:fx},\ref{eq:fez} we can translate
 Eq.\ref{eq:flc} into invariant variables $z$ (we use index I for
 the intermediate state):

\begin{equation}
   dp\cdot dx  = \cdots = 0.5 [ z^+_I (z^-_b - z^-_a) - z^-_I (z^+_b - z^+_a)
      + z^+_a z^-_b - z^-_a z^+_b ] M^2_0
\label{eq:master}  
\end{equation} 
  and it becomes evident that we can involve the interference 
 by an appropriate choice of the $z$ variables 
 (imposing a restriction on the $dp\cdot dx$ term, 
  see again Eqs.\ref{eq:sup},\ref{eq:supa}).

   Let's take a concrete example: during the fragmentation process, imagine 
 the pion $a$ and the arbitrary hadron system $I$  are already generated
 and the pion $b$ (identical with $a$)  is just about to be generated.
 For the two pions to be correlated, we would require their $dp\cdot dx$ term
 to behave according to the interference term in Eq.~\ref{eq:sup}.
 This represents an additional condition on the choice of $z_b$,
 and there is a possibility of correlations in transverse momentum as well.
    We have checked that our 'weighting' model,
 when applied at simple $q\tilde{q}$ string, does not predict
 any strong correlation in the transverse momentum (rather a small decrease
 of the mean transverse momentum is observed),
 and therefore we just
 keep the random generation of transverse momentum of standard JETSET
 -- for simplicity. Having the transverse momentum of pion $b$ defined,
 Eq.~\ref{eq:master} becomes

\begin{equation}
   dp\cdot dx  =  0.5 [ z^+_I (\frac{m^2_{tb}}{z^+_b} -
          \frac{m^2_{ta}}{z^+_a} ) - \frac{m^2_{tI}}{z^+_I} (z^+_b - z^+_a)
      + z^+_a \frac{m^2_{tb}}{z^+_b} - \frac{m^2_{ta}}{z^+_a} z^+_b ]
\end{equation} 

  Obviously, to keep $dp\cdot dx$ at the Planck scale, $z_b$ should be
 close to $z_a$. As a test, we have made a rather simple toy model 
 for coherent fragmentation of simple $u\tilde{u} (d\tilde{d})$ strings.
 We have allowed only charged pions to be produced in the fragmentation, 
 and we have included correlations by a simple rule (keeping notation of Fig.
 \ref{fig:flav}, where I -- intermediate state -- is now represented by
 a pion of the opposite charge):

\begin{itemize}
\item{Case A:} $ z_b^+ = z_a^+ $ if resulting
   $dp\cdot dx  \leq \hbar $      \\
  .... $  dp\cdot dx = 0.5*(1+z_I^+/z_a^+)(m^2_{tb}-m^2_{ta}) $ in this
 scenario 
\item{Case B:} $ z_b^+ = \frac{m_{tb}}{m_{ta}} z_a^+ $
                   if resulting $dp\cdot dx \leq \hbar $   \\
  .... $  dp\cdot dx = 0.5*(m_{tb}/m_{ta}-1)
     (z_I^+z_a^--z_I^-z_a^+) M^2_0 $ in this scenario 
\end{itemize}
 
  ( Actually, we have made the string breaking in flavour $d$ follow
  the pattern  of string breaking in flavour $u$ and vice versa.) 

\begin{figure}[bth]
\mbox{\epsfig{file=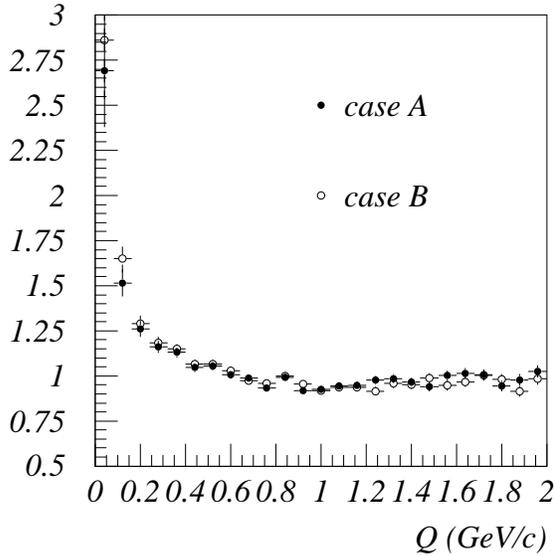,height=10cm,width=10cm}}
\parbox{10.cm}{\caption{\protect\small \sl
 The 2-particle correlation function for equally charged pions in the toy model
(see text).
} 
\label{fig:befr}}
\end{figure}

  As expected, we have obtained a nice enhancement of the two-particle
 correlation function in both cases
 (Fig.\ref{fig:befr}). The correlation is slightly stronger in case B
 since there the difference in transverse momentum is partially compensated.
  A small drop in the mean multiplicity is observed: -4\% in case A,
  -2\% in case B (and consequently
 somewhat harder spectrum of final particles), which we think is
  due to the outside-in method of the fragmentation 
 (when the string is fragmented from both its ends). Actually, the
 inside-out cascade would suit better the implementation of the BE interference
 but the discussion of possible solutions goes beyond the scope of this paper.
   
  The direct implementation of BE interference has many advantages:
  we expect the algorithm to go smoothly over gluon
  corners, and we stress that there is no need to evaluate 
  the absolute coordinates of string breaking and hadron position, 
  which means extreme simplification with respect to our current simulation.
  We also think that a consistent way to treat the short-lived resonances can
  be developed. From practical point of view, the high efficiency of
  simulation, combined with the solid theoretical basis, would considerably
  simplify the study of the BE effect in $e^+e^-$ annihilations.

\section*{Acknowledgements}

   One of us (\v{S}.T.) would like to thank G.Gustafson, R.M{\o}ller,
 T.Sj\"{o}strand and A.Tomaradze for their interest
 to this work and for useful discussions, and D.Bloch, T.Todorov and M.Winter
 for the help with the corrections of this text.  
  
\section*{\appendix{Appendix A}}

  The string area spanned by a simple $q\tilde{q}$ string, parallel to the
 axis $x$ in its rest frame, can be expressed in the terms of the coordinates
 of string breakings. 

\begin{figure}[bth]
\mbox{\epsfig{file=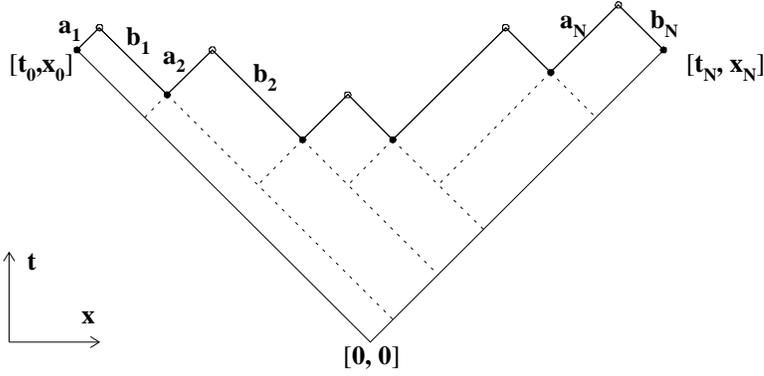,height=6cm,width=12cm}}
\parbox{12.cm}{\caption{\protect\small \sl
  Graphical representation of the calculation of the string area.
} 
\label{fig:app}}
\end{figure}

 In Fig.\ref{fig:app}, string breakings are represented by the full circles
 with coordinates $[t_i,x_i]$, while the empty circles mark the points  where
 the quark-antiquark pairs forming the hadrons meet. If we consider the quarks 
 to be massless, the coordinates of their 'meeting' point are

\begin{equation}
  [t_M,x_M]_i = 0.5 [x_i-x_{i-1}+t_i+t_{i-1}, x_i+x_{i-1}+t_i-t_{i-1}]
\end{equation}

  We introduce variables $a_i, b_i $ (i=1..N) in  the following way

\begin{equation}
   a_i = \sqrt{2} (t_{Mi} - t_{i-1}) = \frac{1}{\sqrt{2}} 
 (x_i-x_{i-1}+t_i-t_{i-1})
\end{equation}

\begin{equation}
   b_i = \sqrt{2} (t_{Mi} - t_{i}) = \frac{1}{\sqrt{2}} 
 (x_i-x_{i-1}-t_i+t_{i-1})
\end{equation}
 ( $a_i,b_i$ are closely related to $z^+_i,z^-_i$ of Eq.\ref{eq:fez}: 
 $ a_i = z^+_i E_{q0}/{\kappa} $;  $ b_i = z^-_i E_{q0}/{\kappa} $ ,
  $E_{q0}$ is the initial energy of endpoint quarks.)  

 The string area may be then written as

\begin{equation}
   A = \sum_{i=1}^{N} a_i b_i + \sum_{i=1}^{N-1} a_i ( b_0 - \sum_{j=1}^{i}
      b_j )
\end{equation}
  where $ b_0 = \sqrt{2} t_0 = \sqrt{2} x_0 = \sqrt{2} E_{q0}/\kappa $ . 

 The string area difference corresponding to an exchange of two
 hadrons ($k,l,k<l$) can be calculated from the previous equation,
 and with the help of Eqs.\ref{eq:f1},\ref{eq:f2} we obtain Eq.\ref{eq:supa} 
 ( [$\tau_i,\chi_i$] are coordinates of the production vertex of
  the hadron i):  

\begin{eqnarray}
  \Delta A & =  & A - A(k\leftrightarrow l)
           = \cdots = ( a_k - a_l ) \sum_{i=k+1}^{l} b_i
              - ( b_k - b_l ) \sum_{i=k+1}^{l} a_i  \nonumber \\
     & =  & 0.5 (x_k-x_{k-1}+t_k-t_{k-1}-x_l+x_{l-1}-t_l+t_{l-1})
  (x_l-x_k-t_l+t_k) \nonumber \\  
     &    & -0.5 (x_k-x_{k-1}-t_k+t_{k-1}-x_l+x_{l-1}+t_l-t_{l-1})
  (x_l-x_k+t_l-t_k) \nonumber \\
     & =  & 0.5 (x_l-x_k-t_l+t_k)(E_k+p_k-E_l-p_l)/\kappa    \nonumber \\   
     &    & -0.5 (x_l-x_k+t_l-t_k)(E_k-p_k-E_l+p_l)/\kappa    \nonumber \\
     & =  & [(E_k-E_l)(t_k-t_l)-(p_k-p_l)(x_k-x_l)]/\kappa \nonumber \\
     & =  & 0.5\{(E_k-E_l) [(t_k+t_{k-1})+(t_k-t_{k-1}) 
                      -(t_l+t_{l-1})+(t_l-t_{l-1})] \nonumber \\
     &    &   - (p_k-p_l) [(x_k+x_{k-1})+(x_k-x_{k-1})
                     -(x_l+x_{l-1})+(x_l-x_{l-1})]\}/\kappa \nonumber \\
     & =  & \{(E_k-E_l) [\tau_k-\tau_l+(p_k-p_l)/2\kappa] 
            - (p_k-p_l) [\chi_k-\chi_l+(E_k-E_l)/2\kappa]\}/\kappa \nonumber \\
     & =  & \{(E_k-E_l)(\tau_k-\tau_l)-(p_k-p_l)(\chi_k-\chi_l)\}/\kappa 
\end{eqnarray}


\begin{thebibliography}{99}

\bibitem{LS}L.L\"{o}nnblad, T.Sj\"{o}strand: {\it Bose-Einstein Effects and 
W Mass Determinations}, CERN-TH/95-17.

\bibitem{JETSET}T.Sj\"{o}strand: {\it PYTHIA 5.7 and JETSET 7.4},
 Computer Physics Commun.82(1994)74.

\bibitem{YR} {\it Physics at LEP200}, CERN Yellow Report, CERN 96-01.

\bibitem{RM}V.Kartvelishvili, R.Kvatadze, R.M{\o}ller: {\it Estimating
 the effects of Bose-Einstein correlations on the W mass measurement at LEP2},
 MC-TH-97/04, MAN/HEP/97/1.

\bibitem{JZ}S.Jadach, K.Zalewski: {\it W mass reconstruction from hadronic 
  events in LEP2-Bose-Einstein Effect}, CERN-TH/97-29.

\bibitem{AH}B.Andersson, W.Hoffmann: {\it Bose-Einstein Correlations and Color
 Strings}, Phys.Letters,169B(1986)364-368.

\bibitem{AB}X.Artru, M.G.Bowler: {\it Quantisation of the string fragmentation
 model}, Z.Phys.C37,293-304(1988).

\bibitem{markus}B.Andersson, M.Ringn\'{e}r: {\it Bose-Einstein Correlations in
 the Lund Model}, LU TP 97-07.


\bibitem{LUN83}B.Andersson, G.Gustafson, G.Ingelman, T.Sj\"{o}strand: {\it
 Parton Fragmentation and String Dynamics}, 
 X.Artru:{\it Classical String Phenomenology. How Strings Work,}\\
 Physics Reports 97,Nos.2\&3(1983)31-171.


\bibitem{tun}DELPHI: {\it Tuning and test of fragmentation models based on
 identified particles and precision event shape data}, Z.Phys.C73,11-59(1996).

\bibitem{BE-DELPHI}DELPHI Coll.,CERN/PPE 94-02 and Z.Phys.C63(1994)17, \\
                and A.Tomaradze, private communication.


\bibitem{rho}OPAL Coll.,Z.Phys.C56(1992)521;\\
             DELPHI Coll.,Z.Phys.C65(1995)587;\\
             ALEPH Coll.,Z.Phys.C69(1996)379.  

\bibitem {BiaPe} A. Bia{\l}as and R. Peschanski, Nucl.Phys.B273 (1986) 703; 
Nucl. Phys. B308 (1986) 857.

\bibitem {ptilda} A. Bia{\l}as and M. Gazdzicky, Phys.Lett.B252 (1990) 483.; \\
 W. Ochs, Z. Phys. C50 (1991) 339.

\bibitem {BiaVietri} A.Bia{\l}as, in MULTIPARTICLE DYNAMICS 1994, ed. A.
Giovannini, S.Lupia and R. Ugoccioni, World Scientific, 1995.

\bibitem {AmiDEL} P. Abreau et al. (DELPHI Coll.), Z. Phys.C63 (1994) 17.

\bibitem {DeWolf} E.A. De Wolf, I.M. Dremin and W. Kittel, Phys.Rep.270 (1996)
 1.                 

\bibitem{CR}T.Sj\"{o}strand, V.A.Khoze: {\it On Colour Rearrangement 
in Hadronic WW Events}, Z.Phys.C62(1994)281-309;\\
            L.L\"{o}nnblad:{\it Reconnecting Colour Dipoles},
  CERN-TH/95-218.  


\bibitem{me}\v{S}.Todorova-Nov\'a: {\it Colour Reconnection in String Model},
     DELPHI note 96-158 (public).
 

\bibitem{wwmes}DELPHI Coll.: {\it Measurement of Correlations between
 Pions from Different W's in $e^+e^->W^+W^-$ Events },
 CERN-PPE/97-30 and Phys.Lett.B401(1997)181.



\end{thebibliography}
\end{document}